\documentclass[9pt,onecolumn,twoside]{arxiv}
\usepackage[]{geometry}
\usepackage[scriptsize,tight]{subfigure}
\usepackage{graphicx}
\usepackage{cite}
\usepackage{multirow}
\usepackage{color,soul}
\usepackage{hyperref}

\articletype{inv} 

\runningtitle{pH-Responsive Glyphosate Adsorption on Hydroxylated Carbon Nanotubes}

\runningauthor{Silva \textit{et al.}}

\title{pH-Responsive Glyphosate Adsorption on Hydroxylated Carbon Nanotubes: From Electronic Structure to Molecular Dynamics}

\author[1]{H.~T.~Silva}
\author[1]{L.~C.~S.~Faria}
\author[2]{T.~A.~Aversi-Ferreira}
\author[1,$\ast$]{I.~Camps}
\correspondingauthoraffiliation[$\ast$]{icamps@unifal-mg.edu.br}

\affil[1]{Laborat\'orio de Modelagem Computacional - \emph{La}Model,
Instituto de Ci\^{e}ncias Exatas - ICEx. Universidade Federal de Alfenas -
UNIFAL-MG, Alfenas, Minas Gerais, Brazil}

\affil[2]{Laboratory of Biomathematics, Institute of Science and Tecnology - ICT, Federal University of Alfenas - UNIFAL-MG, Poços de Caldas, Minas Gerais, Brazil}

\begin{abstract}
This computational study investigates glyphosate adsorption mechanisms on hy\-drox\-yl-func\-tion\-al\-ized carbon nanotubes (CNTs) as an alternative approach for environmental remediation. Single-walled CNTs with (10,0) zigzag chirality were functionalized with hydroxyl groups at concentrations of 5-25\% and evaluated for interactions with glyphosate in five different ionization states (G1-G5) corresponding to pH-dependent protonation. Using semi-empirical tight-binding methods implemented in xTB software, molecular geometry optimization, electronic property calculations, topological analyses via Quantum Theory of Atoms in Molecules (QTAIM), and molecular dynamics simulations at 300~K were performed. Results demonstrate that functionalization significantly enhances adsorption capacity, with binding energies becoming increasingly negative at higher OH concentrations and with more deprotonated glyphosate forms (G4 and G5). Electronic coupling analyses reveal optimized charge reactivity and transport in systems with 20-25\% OH functionalization. Topological characterization identified 477 bond critical points, confirming donor-acceptor interactions with strong covalent contributions, particularly in highly functionalized systems. Radial distribution function profiles from molecular dynamics simulations demonstrate that functionalization promotes spatial organization on nanotube surfaces, increasing contact regions and reducing molecular mobility. Systems with moderate interactions (CNT+OH\textsubscript{x}+G1 and CNT+OH\textsubscript{x}+G3) present environmentally and economically viable solutions, enabling adsorbent regeneration and reuse. The findings indicate that OH-functionalized carbon nanotubes show significant promise for glyphosate detection and capture applications in environmental monitoring and remediation, regardless of the pesticide's ionization state.
\end{abstract}


\begin{document}
\maketitle
\thispagestyle{firststyle}
\vspace{-13pt}

\section{INTRODUCTION}
\label{Sec:Intro}
The use of agrochemicals in agriculture, focusing on large-scale production, began after World War II, when new compounds were synthesized for military purposes~\cite{01_thomine_using_2022}. Glyphosate [N-(phosphonomethyl) glycine], a synthetic, non-selective compound widely commercialized for controlling weeds and invasive plantation species, became one of the most used active ingredients globally~\cite{03_meftaul_controversies_2020,04_leoci_glyphosate_2020,02_martins-gomes_glyphosate_2022}. Its half-life can be long, with significant environmental persistence ranging from 0.8 to 151 days, presenting risks to fauna and flora. Moreover, exposure to glyphosate (GLY) and its byproducts like aminomethylphosphonic acid (AMPA), even in small concentrations, affects human health and ecosystem stability~\cite{06_diel_carbon_2021,05_rivas-garcia_overview_2022}.

The intensive and indiscriminate use of this agrochemical has caused its accumulation in various environmental compartments, compromising air, water, and soil quality. This leads to adverse consequences for biota and human health, contributing to the development and aggravation of diseases such as cancer, diabetes, and exhibiting carcinogenic, mutagenic, genotoxic, neurological, reproductive, and teratogenic effects~\cite{07_kier_review_2013,08_genotoxicity_2022}. Numerous studies have documented impacts on fauna, including elimination of essential biotic system organisms, behavioral alterations in bees, aneurysms, epithelial cell hyperplasia, capillary changes in fish, and DNA damage in amphibians~\cite{10_goncalves_detecting_2015,09_campos-garcia_histopathological_2016}.

It has become desirable not only to reduce applied glyphosate quantities for greater prevention but also to ensure its efficient environmental removal~\cite{11_ampilek_17_2020}. Various technologies like membrane separation methods, electrolysis, photocatalytic degradation, advanced oxidative processes, microwave radiation, ozonation, and ultraviolet irradiation have been applied to remove glyphosate from different sample types. However, these methods have proven insufficient, presenting limited flexibility, high costs, low efficiency, and potential secondary pollutant production~\cite{13_yang_simultaneous_2018,12_gaberell_highly_2019,06_diel_carbon_2021}.

Nanotechnology is considered an emerging field providing an alternative and most appropriate technology for efficiently removing pollutants at low operational costs~\cite{14_thirunavukkarasu_review_2020}. It has become a powerful tool for removing toxic agents from the environment, as its minute size provides a larger surface area, subsequently increasing reactive surface potential. Nanomaterials also possess a unique surface chemistry, allowing functionalization or grafting with functional groups to adsorb agrochemicals efficiently\cite{15_sebastian_chapter_2020}.

In this context, carbon nanotubes when functionalized with carboxyl, hydroxyl, phenolic, and amine groups possess interesting characteristics like nanocapillarity, porosity, extensive surface area, and environment-sensitive electronic properties. These features enable efficient adsorption of glyphosate or its degradation products~\cite{18_rahman_overview_2019,16_arora_carbon_2020,17_aligayev_chapter_2022}. Additionally, they demonstrate adsorbent regeneration and reuse capabilities, enabling long-term process viability~\cite{11_ampilek_17_2020,06_diel_carbon_2021}.

Due to carbon atoms strong covalent bonds on carbon nanotube surfaces, these systems are highly stable with considerable chemical inertness. Many applications of these structures require chemical modification to adjust and control their physicochemical properties~\cite{19_chemistry_2006}. Exploring carbon nanotubes potential involves functionalization through walls, tips, or encapsulation by binding functional groups (-OH, -COOH, -NH\textsubscript{2}) and atoms to the nanotube wall, doping, producing defects (vacancies), and deformation~\cite{23_milowska_functionalization_2013,20_cardenas-benitez_reviewcovalent_2018,21_fiyadh_review_2019,22_Salah_2021}.

This procedure allows modification of electronic and mechanical properties, suggesting increased material reactivity, enhancing cationic exchange and electrostatic interactions~\cite{24_mohd_nurazzi_fabrication_2021}. This facilitates nanotubes interaction with pharmaceuticals, organic, biological, or toxic molecules~\cite{25_oliveira_nanotubos_2011}. Through semi-empirical tight binding studies, this research explored glyphosate adsorption mechanisms on functionalized carbon nanotube surfaces, aiming to provide molecular-level insights into these interactions and offer a novel approach to mitigating glyphosate pollution.

\section{MATERIALS AND METHODS}
\label{Sec:Methods}
Single-walled carbon nanotubes with a (10,0) chirality, zigzag configuration, semiconductor character, and dimensions of 7.83~{\AA} in diameter and 12.78~{\AA} in length were used in both pure and hydroxyl-functionalized states at concentrations of 5, 10, 15, 20, and 25\%. For each concentration, the representative structure was selected based on the highest system entropy, employing quasi-entropy as the selection criterion~\cite{26_ribeiro_effects_2017}.

The interaction analysis between glyphosate and functionalized carbon nanotubes was conducted using the semi-empirical tight-binding method implemented in the xTB (extended tight-binding) software package, which is self-consistent, precise, and includes electrostatic multipole contributions and density-dependent dispersion~\cite{27_bannwarth_extended_2021}. The molecular geometry was optimized to find the configuration with the lowest potential energy by adjusting atomic positions. Atomic coordinates were systematically modified until the calculated energy reached its lowest value and achieved its fundamental state~\cite{28_schlegel_geometry_2011}.

Structure optimization was performed at an extreme level, with an energy convergence of $5\times10^{-5}\,E_h$ and gradient norm convergence of $5\times10^{-5}\,E_h/a_0$ (where $a_0$ is the Bohr radius)~\cite{29_aguiar_exploring_2024}.

\begin{figure}[tbph]
\centering
\includegraphics[width=10cm]{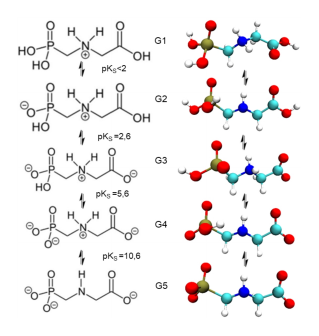}
\caption{\label{Fig:Gly} 2D and 3D illustration of the ionic glyphosate forms. Representations based on pKa values associated with each protonation state.}
\end{figure}

The formed complexes involved the hydroxyl functional group (OH\textsubscript{x}), where x represents concentrations of 0, 5, 10, 15, 20, and 25\%, and glyphosate in different ionization states. These states were defined based on glyphosate's acid dissociation constants (pKa: 2.0, 2.6, 5.6, and 10.6), which are determined by the pH of the surrounding medium: G1 ($pH < 2$), G2 ($pH \approx 2-3$), G3 ($pH \approx 4-6$), G4 ($pH \approx 7-10$), and G5 ($pH > 10.6$), as shown in Figure~\ref{Fig:Gly}~\cite{30_dissanayake_herath_statistical_2019}. The nomenclature used to identify the systems was defined as CNT+OH\textsubscript{x}+Gy, where CNT refers to carbon nanotubes, OH\textsubscript{x} to the hydroxyl functional group, and Gy represents the specific ionized form of glyphosate considered in each complex.

\begin{figure}[tbph]
\centering
\includegraphics[width=15cm]{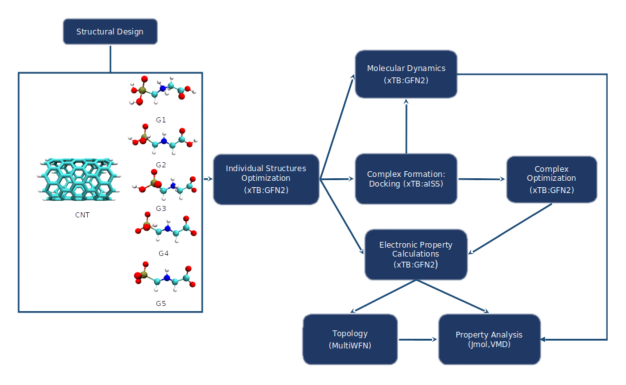}
\caption{\label{Fig:Methods} Computational procedure flowchart.}
\end{figure}

The calculations were performed according to the procedure sequence presented in the flowchart of Figure~\ref{Fig:Methods}. Initially, the geometries of isolated structures of nanotubes and glyphosate were optimized. Subsequently, a coupling process was conducted using automated interaction site mapping (aISS). This approach sought to identify accessible regions on the nanotube surfaces (molecule A), followed by a three-dimensional (3D) screening to identify $\pi-\pi$ type interactions in various directions. Afterward, adjustments were made to identify the most favorable orientations of molecule B (glyphosate) around molecule A (nanotubes)~\cite{29_aguiar_exploring_2024}.

To classify the generated structures, interaction energy was used. A default two-step refinement protocol based on genetic algorithms was adopted, where one hundred structures with the lowest interaction energies were selected to ensure that conformations not detected in the initial screening were included. During this two-stage genetic optimization procedure, each pair of glyphosate molecule positions was randomly combined around the carbon nanotubes. Subsequently, 50\% of the structures underwent random mutations in both position and angle. After ten iterations of this search process, ten complexes with the lowest interaction energies were selected. Thus, the structure with the lowest interaction energy was chosen as the input for complex optimization~\cite{29_aguiar_exploring_2024}.

Electronic properties were determined using the spin polarization scheme, and the following parameters were calculated for the systems: HOMO ($\varepsilon_H$); LUMO ($\varepsilon_L$); energy gap between HOMO and LUMO orbitals ($\Delta \varepsilon = \varepsilon_H - \varepsilon_L$); $J_{oc}$ (Occupied Orbital Interaction Energy); $J_{un}$ (Unoccupied Orbital Interaction Energy); $J$ (Transition Orbital Interaction Energy), total energy, and dispersion energy.

To investigate charge carrier mobility and evaluate electronic interactions between nanotubes and glyphosate molecules, electronic transfer integrals were calculated using the dimer projection method (DIPRO). In the analyses, $J_{oc}$ denotes hole transport (occupied molecular orbitals) and high values indicate greater decoupling between two fragments, signifying stronger repulsion or stabilization effects between filled orbitals. $J_{un}$ represents electron transport (unoccupied molecular orbitals), and high values suggest strong orbital interactions that facilitate electronic transitions. $J$ represents total charge transfer, encompassing both hole and electron transport between occupied and unoccupied molecular orbitals, respectively. Higher $J$ values indicate chemical reactivity and changes in the molecule's electronic state with strong coupling between two fragments~\cite{31_albright_orbital_2013,32_rauk_orbital_2004,33_kohn_efficient_2023}.

The adsorption energy ($E_{ads}$) is calculated as the difference between the energy of the final system CNT+OH\textsubscript{x}+Gy ($E_{CNT+OH\textsubscript{x}+Gy}$) and the sum of initial isolated carbon nanotubes ($E_{CNT+OH\textsubscript{x}}$) and glyphosate system ($E_{Gy}$) energies:

\begin{equation}
\label{Eq:bind}
E_{ads} = E_{CNT+OH\textsubscript{x}+Gy} - E_{CNT+OH\textsubscript{x}}- E_{Gy}.
\end{equation}

To understand and classify the types and strength of interactions formed between glyphosate and carbon nanotubes, a topological properties study was conducted using the wave function obtained from electronic property calculations (29). This approach allowed identification of bond critical points (BCPs) and quantification of descriptors such as electronic density ($\rho$), Laplacian ($\nabla^2 \rho$), electron localization function (ELF), and localized orbital locator (LOL), which were analyzed using the MULTIWFN software utilizing the wave function generated during electronic property calculations~\cite{34_lu_multiwfn_2012,Multiwfn2}.

To evaluate the strength and bond type between attractive atom pairs, only \textbf{(3,-1)} bond critical points were analyzed, as these are characterized by minimal electronic density along the bond path between two nuclei at the glyphosate molecule-nanotube interface. By doing so, the researchers obtained parameters based on a physical observable (electronic density) that is free from bias, complementing wave function or molecular orbital analysis techniques. This method avoids attributing physical meaning to a specific orbital set, preventing the potential loss of important details that might occur when relying solely on orbital-based analyses. Moreover, electronic density offers the advantage of being analyzable through both theoretical and experimental approaches~\cite{36_bader_atoms_1994,35_Koch_2024,37_fedorov_topological_2025}.

In contrast to geometric optimization, which seeks to identify the lowest-energy structure on a potential energy surface, molecular dynamics (MD) simulations allowed examination of glyphosate molecule movements, providing a more comprehensive understanding of the system's dynamic behavior~\cite{38_martinez_packing_2003}. The molecular dynamics simulations were conducted at 300~K with a production run time of 100~ps, using a time step of 2~fs and a dump step of 50~fs, with the final configuration recorded in a trajectory file. These calculations employed the GFN-FF force field, specifically designed to balance high computational efficiency with the precision typically associated with quantum mechanics methods.

To characterize the spatial distribution of glyphosate molecules, the radial distribution function (RDF) was employed:

\begin{equation}
\label{Eq:RDF}
g(\mathbf{r}) = \frac{n(\mathbf{r})}{4 \pi \rho \mathbf{r}^2 \Delta \mathbf{r}},
\end{equation}
where $n(\mathbf{r})$ is the mean number of particles in a shell of width $\Delta \mathbf{r}$ at distance $\mathbf{r}$, and $\rho$ is the mean particle density.

Statistically, $g(\mathbf{r})$ describes the probability of finding a glyphosate molecule at a position $\mathbf{r}$ relative to the carbon nanotube, normalized by the average density. This analysis proves particularly useful in heterogeneous systems like the one studied, as it helps predict how glyphosate molecules organize themselves in relation to the nanotube surface~\cite{39_hansen_theory_2013}.

\section{RESULTS AND DISCUSSION}
\label{Sec:Results}
\subsection{Electronic properties}
\label{Sec:Struct-Elect}
\begin{figure}[tbph]
\centering
\includegraphics[width=10cm]{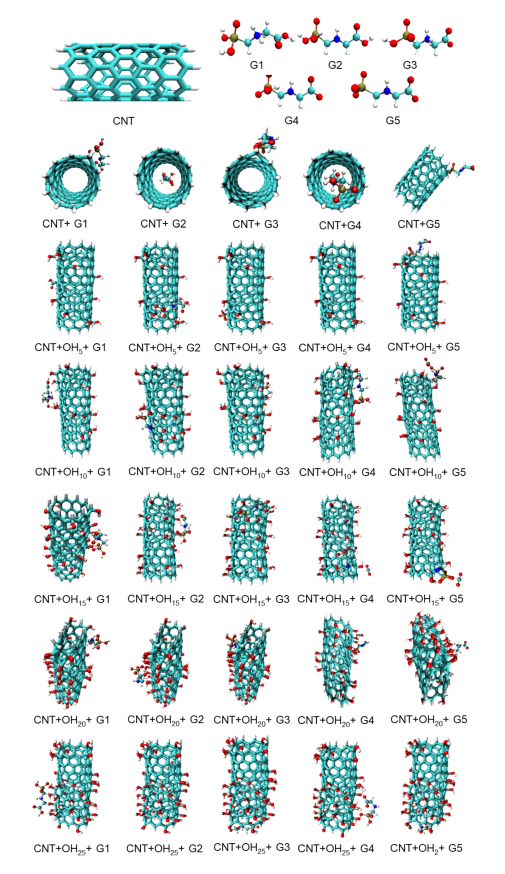}
\caption{\label{Fig:OptGeo} Optimized structures for pristine CNT, functionalized nanotubes, glyphosate and complexes.}
\end{figure}

Figure~\ref{Fig:OptGeo} presents the conformations of complexes formed between glyphosate and carbon nanotubes. This geometric optimization stage is crucial in computational adsorption studies, as it allows identification of the structural configuration with the lowest potential energy-essentially the fundamental state-and helps understand molecular interaction mechanisms. Moreover, it ensures that the obtained structure represents an energetically stable configuration for subsequent electronic, thermodynamic, and kinetic property calculations used in molecular reactivity and interaction analyses~\cite{28_schlegel_geometry_2011,40_yang_machine-learning_2021}.

\begin{figure}[tbph]
\centering
\includegraphics[width=10cm]{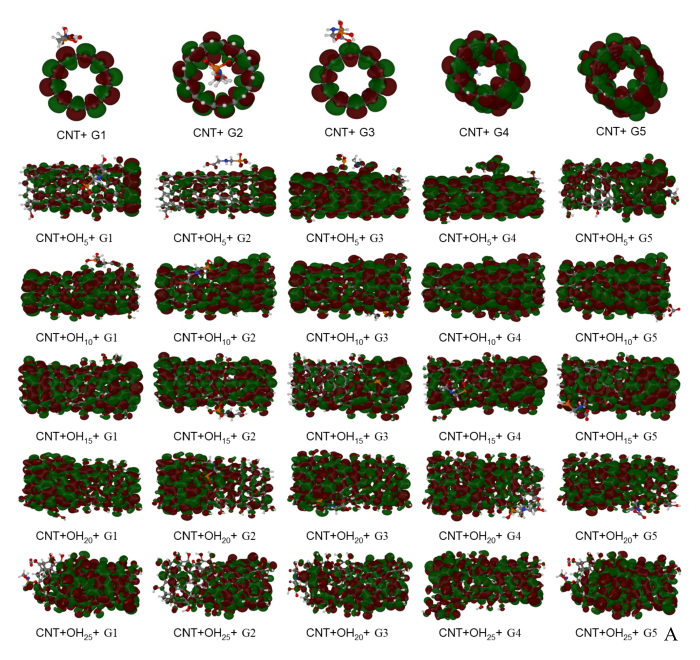}
\includegraphics[width=10cm]{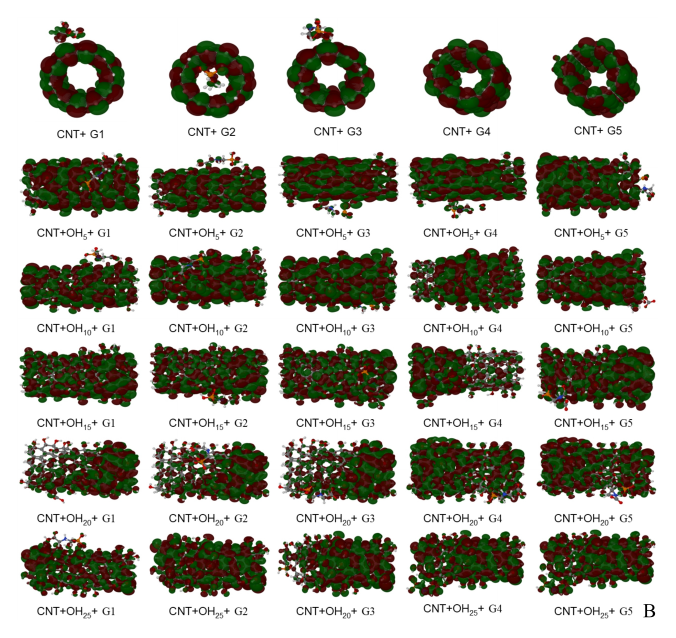}
\caption{\label{Fig:Orbitals} Structure of the frontiers orbitals of all systems. (A) HOMO and (B) LUMO.}
\end{figure}

Structural changes between carbon nanotube (CNT) and glyphosate geometries, shown in Figure~\ref{Fig:Orbitals}, become more pronounced depending on the glyphosate molecule's state and the concentration of OH groups used in nanotube functionalization. For functionalized nanotube systems, a higher degree of structural distortion is observed compared to pure CNTs, due to their increased charge density and high hydrogen bonding capacity resulting from OH group presence~\cite{41_GaddielSandoval-ACSMaterialsAu-2025}.

Functionalization effects on geometry correlate with glyphosate species nature. In G1 and G2 conformations with higher protonation, the global charge and negative charge density of active groups are smaller, potentially promoting interactions closer to nanotube edges. Conversely, G4 and G5 are more deprotonated with negative charges, thus tending to induce pronounced rearrangements in CNT orientation and curvature~\cite{42_lara_functionalization_2014}. In this context, Ribeiro \emph{et al.}~\cite{26_ribeiro_effects_2017} affirms that this structural response is compatible with CNTs' sensitivity, as their hexagonal structure with $sp^2$ hybridization is highly susceptible to surface perturbations that can locally transform the carbon atom's electronic character to $sp^3$. The presence of ionized glyphosate and OH functionalization can impact properties like electrical conductivity, state density, and electronic affinity, potentially affecting system geometry by partially disrupting carbon ring symmetry and causing structural tensions~\cite{23_milowska_functionalization_2013,43_meunier_physical_2016,44_vargas-delgadillo_nanostructured_2022}. Furthermore, in systems with high OH concentrations like CNT+OH\textsubscript{20}+Gy and CNT+OH\textsubscript{25}+Gy, multiple OH groups coexisting in close proximity can generate tension accumulation and propagate curvatures or distortions to adjacent regions~\cite{23_milowska_functionalization_2013}.

Table~\ref{Tab:ElectProp} presents electronic and reactivity data (binding energy, HOMO and LUMO orbital energies, electronic gap ($\Delta \varepsilon$), and electronic coupling parameters ($J_{oc}$, $J_{un}$, $J$)) for complexes formed between carbon nanotubes and glyphosate. These indicators were analyzed to understand system stability and charge transfer potential, directly influencing adsorption selectivity and efficiency.

\begin{table}[tbph]
\caption{Adsorption energy and electronic properties$^\dagger$.}
\label{Tab:ElectProp}
\begin{center}
\setlength\extrarowheight{-3pt}
\begin{tabular}{lrrrrrrr}
\hline
System      & $E_{ads}$ & $\varepsilon_H$ & $\varepsilon_L$ & $\Delta \varepsilon$ & $J_{oc}$ & $J_{un}$ & $J$   \\
\hline
\hline
CNT         & 0         & -8.86           & -8.86           & 0.00                 & -        & -        & -     \\
CNT+G1      & -3.27     & -11.08          & -11.04          & 0.04                 & 0.01     & 0.01     & 0.01  \\
CNT+G2      & -1.02     & -8.90           & -8.90           & 0.00                 & 0.02     & 0.01     & 0.018 \\
CNT+G3      & -2.44     & -6.65           & -6.60           & 0.05                 & -        & -        & -     \\
CNT+G4      & -4.49     & -5.12           & -5.08           & 0.04                 & -        & -        & -     \\
CNT+G5      & -9.84     & -3.40           & -3.36           & 0.04                 & -        & -        & -     \\
CNT+OH05+G1 & -2.27     & -10.71          & -10.71          & 0.00                 & 0.07     & 0.00     & 0.02  \\
CNT+OH05+G2 & -1.22     & -9.22           & -9.21           & 0.01                 & 0.01     & 0.02     & 0.02  \\
CNT+OH05+G3 & -1.67     & -7.72           & -7.72           & 0.01                 & 0.02     & 0.02     & 0.02  \\
CNT+OH05+G4 & -7.01     & -5.97           & -5.96           & 0.01                 & 0.01     & 0.01     & 0.03  \\
CNT+OH05+G5 & -11.69    & -4.01           & -3.97           & 0.04                 & 0.01     & 0.03     & 0.02  \\
CNT+OH10+G1 & -3.76     & -11.06          & -11.02          & 0.05                 & 0.03     & 0.02     & 0.02  \\
CNT+OH10+G2 & -1.44     & -9.20           & -9.18           & 0.02                 & 0.01     & 0.00     & 0.01  \\
CNT+OH10+G3 & -2.38     & -7.70           & -7.67           & 0.03                 & 0.02     & 0.04     & 0.02  \\
CNT+OH10+G4 & -5.56     & -6.00           & -5.98           & 0.02                 & 0.00     & 0.00     & 0.01  \\
CNT+OH10+G5 & -13.05    & -4.36           & -4.35           & 0.01                 & 0.00     & 0.00     & 0.01  \\
CNT+OH15+G1 & -3.52     & -11.18          & -11.16          & 0.02                 & 0.00     & 0.00     & 0.01  \\
CNT+OH15+G2 & -1.60     & -9.44           & -9.44           & 0.01                 & 0.02     & 0.01     & 0.02  \\
CNT+OH15+G3 & -3.47     & -7.91           & -7.88           & 0.03                 & 0.01     & 0.01     & 0.04  \\
CNT+OH15+G4 & -7.01     & -6.21           & -6.20           & 0.02                 & 0.02     & 0.06     & 0.05  \\
CNT+OH15+G5 & -14.92    & -4.67           & -4.64           & 0.03                 & -        & -        & -     \\
CNT+OH20+G1 & -3.13     & -11.36          & -11.33          & 0.03                 & 0.00     & 0.01     & 0.01  \\
CNT+OH20+G2 & -1.73     & -9.57           & -9.56           & 0.01                 & 0.02     & 0.01     & 0.02  \\
CNT+OH20+G3 & -2.38     & -8.02           & -8.02           & 0.01                 & 0.02     & 0.00     & 0.01  \\
CNT+OH20+G4 & -7.58     & -5.93           & -5.90           & 0.03                 & 0.02     & 0.02     & 0.17  \\
CNT+OH20+G5 & -16.06    & -4.60           & -4.59           & 0.01                 & 0.01     & 0.03     & 0.12  \\
CNT+OH25+G1 & -4.83     & -9.30           & -10.76          & 0.02                 & 0.00     & 0.00     & 0.01  \\
CNT+OH25+G2 & -1.19     & -9.30           & -9.29           & 0.01                 & 0.01     & 0.07     & 0.04  \\
CNT+OH25+G3 & -3.40     & -7.73           & -7.72           & 0.01                 & 0.08     & 0.03     & 0.05  \\
CNT+OH25+G4 & -7.67     & -6.18           & -6.16           & 0.02                 & 0.02     & 0.00     & 0.01  \\
CNT+OH25+G5 & -16.88    & -4.41           & -4.38           & 0.03                 & 0.00     & 0.00     & 0.01  \\
\hline
\end{tabular}
\begin{flushleft}
\tiny {$^\dagger$ Electronic properties were calculated using an electronic temperature of $300\,K$. Subsequently, the HOMO and LUMO energy levels were determined using the Fermi energy. All energies are in units of eV.}
\end{flushleft}
\end{center}
\end{table}

The binding energy values in Table~\ref{Tab:ElectProp} demonstrate a trend of intensifying interactions between carbon nanotubes and glyphosate as hydroxyl group (OH) concentration increases on CNT surfaces. This is particularly evident in CNT+OH\textsubscript{x}+G5 and CNT+OH\textsubscript{x}+G4, which exhibited more negative binding energy values, indicating greater stability of the formed complexes. This occurs due to higher negative charge density, favoring hydrogen bonding and electrostatic interactions with functional adsorbent sites, thereby modulating surface charge and adsorbate adsorption capacity, promoting enhanced system energy stability~\cite{06_diel_carbon_2021,45_SAraujo-2024}.

For CNT+OH\textsubscript{x}+G1, CNT+OH\textsubscript{x}+G2, and CNT+OH\textsubscript{x}+G3 systems, glyphosate molecule configuration significantly influences binding energy values, which remain moderate without a linear trend. Protonated or neutral forms (G1, G2, and G3) show less negative values, indicating weaker and less stable interactions from an energetic perspective. This aligns with these species' lower charge density and suggests that hydroxyl functionalization promotes interactions between glyphosate and nanotubes, potentially due to increased hydrogen bonding and van der Waals forces, as hydroxyl groups are highly prone to forming hydrogen bonds with organic contaminants~\cite{30_dissanayake_herath_statistical_2019,46_dong_hydrogen_2024}.

The results correspond with previous analysis~\cite{47_ramrakhiani_utilization_2019}, which affirms that pH affects adsorbent surface charge, functional group states, and adsorbate ionization degree-key factors influencing adsorption processes. As shown in Table~\ref{Tab:ElectProp}, systems containing more deprotonated glyphosate forms (G4 and G5) demonstrated the most negative binding energy values, indicating greater stability and affinity between the adsorbate and functionalized surface. Similar behavior was reported in~\cite{06_diel_carbon_2021}, where glyphosate adsorption in aqueous matrix with MWCNT/MPNs-Fe was accentuated under acidic conditions. Lowering solution pH from 10 to 4 caused an increase in herbicide removal percentage, whereas a decrease from pH 4 to 3 led to a slight reduction in these parameters.

Electrostatic attraction and repulsion interactions between functionalized nanotube binding sites and herbicide ionized forms contribute to the Table~\ref{Tab:ElectProp} results. Glyphosate possesses an amphoteric nature and zwitterionic structure, simultaneously presenting positive and negative charges in different molecular regions depending on solution pH. This glyphosate behavior reflects the stability of complexes formed with nanotubes~\cite{30_dissanayake_herath_statistical_2019,47_ramrakhiani_utilization_2019,06_diel_carbon_2021}.

In the dissociation context, glyphosate has equilibrium constants associated with pKa values: 2.0, 2.6, 5.6, and 10.6. Under extremely acidic conditions ($pH < 2.0$), it predominantly carries a positive charge. At intermediate pH levels between 2.0 and 2.6, the molecule appears neutral, while above 2.6, glyphosate's global charge becomes increasingly negative with pH increase. These variations in total glyphosate molecule charge can impact electrostatic interactions with functionalized nanotubes and influence observed binding energies~\cite{30_dissanayake_herath_statistical_2019,06_diel_carbon_2021}.

At low pH, opposing charges between glyphosate and nanotube functional groups may favor more stable interactions, reflected in more negative binding energies. Conversely, at pH levels where glyphosate carries a neutral charge or interacts with similarly charged sites, electrostatic repulsions tend to predominate, resulting in less negative binding energy values and less stable systems.

Adsorbent reutilization viability correlates with regeneration capacity-an important economic and environmental consideration enabling prolonged material usage, reducing costs and minimizing environmental impacts~\cite{49_zavareh_modification_2018,48_krishnamoorthy_date_2019,06_diel_carbon_2021}. Systems with extremely strong interactions like CNT+OH\textsubscript{x}+G4 and CNT+OH\textsubscript{x}+G5 present regeneration and nanotube reuse difficulties due to high binding energy. Conversely, systems with moderate interactions such as CNT+OH\textsubscript{x}+G1 and CNT+OH\textsubscript{x}+G3 allow easier desorption, facilitating material recycling. For CNT+OH\textsubscript{x}+G2, low binding suggests inefficient glyphosate adsorption, potentially leaving the contaminant available in the environment.

Frontier orbital energies for LUMO and HOMO (Figure~\ref{Fig:Orbitals} and Table~\ref{Tab:ElectProp}) were used alongside other electronic properties to assess molecular reactivity, characterizing the molecule's electron donation and acceptance capabilities~\cite{53_guo_mechanical_2003,52_Dalton-ChemRev-2009,51_chandrasekaran_structural_2015,50_reber_superatoms_2017}. The systems exhibit distinct values corresponding to glyphosate ionization levels, with minimal gap variation. The CNT+G2 system possesses the lowest gap value and tends to be the most unstable, consistent with the partially ionized glyphosate form capable of modulating local electronic density. Conversely, CNT+G4, CNT+G3, CNT+G1, and CNT+G5 demonstrate higher gap values, indicating greater molecular stability and consequently lower reactivity in interactions.

For functionalized systems, different glyphosate forms modify HOMO and LUMO values compared to pristine CNTs. The more deprotonated G4 and G5 forms promote an increase in these properties, making them less negative, with reduced electronic stabilization and higher reactivity potential. This may be associated with introducing electron-donating centers via sp\textsuperscript{3} hybridization of carbons linked to hydroxyl groups, potentially creating intermediate electronic states through interactions between glyphosate molecule charges and hydroxyl groups~\cite{54_ghasemi_interatomic_2015,55_speranza_role_2019}.

For G1 and G2 at lower OH concentrations (5\%, 10\%), HOMO and LUMO values remain more negative, suggesting electronic structure stabilization. An atypical behavior is observed in CNT+OH\textsubscript{x}+G5 (5\%) and CNT+OH\textsubscript{x}+G1 (25\%) systems, potentially linked to local conformational rearrangements during geometric optimization. These are influenced by asymmetric charge distributions in highly protonated (G1) or extensively deprotonated (G5) species, as suggested in~\cite{56_zhang_structured_2022,46_dong_hydrogen_2024}.

With hydroxyl group introduction on nanotube surfaces, fluctuations in $\Delta \varepsilon$ values occur, with reductions in some cases, suggesting local electronic density redistribution rather than linear electronic stabilization. This stems from hydroxyl groups' polar nature, where their presence can alter electronic charge distribution, potentially resulting in stabilization or increased reactivity depending on glyphosate's specific form in the system~\cite{58_machado_de_menezes_pristine_2014,57_Deline-ChemRev-2020}. For more protonated G1 and G2 forms, functionalization effects on gap were less pronounced, indicating less effective orbital interaction. This results from lower available charge density for orbital coupling, with minimal electronic reorganization at the nanotube-glyphosate interface, as the system's electron donation capacity varies with its ionized form~\cite{50_reber_superatoms_2017}.

Coupling parameter analysis in ~\ref{Tab:ElectProp} represents charge transfer efficiency between occupied ($J_{oc}$), unoccupied ($J_{un}$) orbitals, and total transfer contribution ($J$), which can be used to evaluate material adsorption capacity~\cite{59_Coropceanu-ChemRev-2007,60_troisi_charge_2011}. Functionalized systems demonstrate elevated values compared to pristine CNTs, showing that OH groups increase charge delocalization and electronic reactivity. Notably, CNT+OH\textsubscript{20}+G4 and CNT+OH\textsubscript{20}+G5 exhibited high $J$ values, indicating substantial electronic transport capabilities. Additionally, in the absence of functionalization, the G2 glyphosate form generated significant $J$ values, demonstrating how herbicide ionization degree influences electronic coupling capacity.

\subsection{Topological Analysis}
\label{Sec:Topo}
Topological analysis based on the Quantum Theory of Atoms in Molecules (QTAIM) provides insights into interactions between carbon nanotubes and glyphosate pesticide, enabling evaluation of bond versatility and nanotube structural stability~\cite{61_okon_single-atom_2023}. By analyzing bond critical points with electronic density topological descriptors like $\rho(\mathbf{r})$ and $\nabla^2\rho(\mathbf{r})$, researchers could identify and classify inter- and intramolecular interactions, whether covalent or non-covalent. Among the latter, these include electrostatic interactions, weak van der Waals forces, and hydrogen bonds-characterizations fundamental to understanding interaction nature and intensity within the system~\cite{62_shahbazian_why_2018,63_agwupuye_electronic_2021,64_yadav_adsorption_2022}.

The simultaneous presence of a bond path and bond critical point between two atoms is a necessary and sufficient condition for chemical interaction, although it is widely recognized that alternative interpretations may occur, especially for weak or very weak interactions~\cite{65_cabeza_topological_2009}. The presence of \textbf{(3,-1)} type bond critical points confirms bonding formation between nanomaterial surfaces and glyphosate, indicating stable interactions that enhance adsorption capacity after functionalization.

\begin{figure}[tbph]
\centering
\includegraphics[width=10cm]{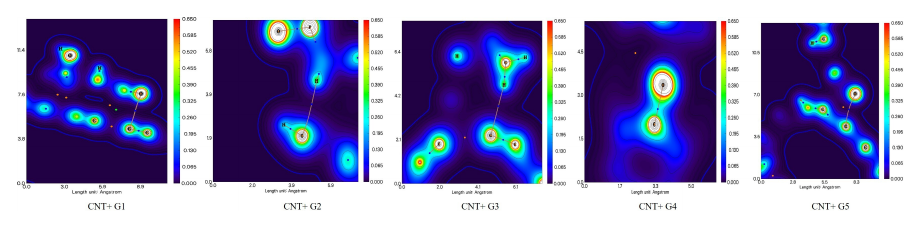}
\caption{\label{Fig:Rho} 2D representation of the electronic density, $\rho$.}
\end{figure}

From the calculations, 477 \textbf{(3,-1)} type critical points were identified, distributed across different systems according to topological descriptor traces and profiles. The bond type (covalent or non-covalent) can be classified based on electronic density ($\rho$) and its Laplacian ($\nabla^2\rho$). Bonds with $\rho > 0.20$~a.u. and $\nabla^2\rho < 0$ are characterized as covalent, while those with $\rho < 0.10$~a.u. and $\nabla^2\rho > 0$ indicate non-covalent interactions. Non-covalent regions are highlighted in orange for each descriptor in Figure~\ref{Fig:Rho}.

Electronic density values at critical points ($\rho$) vary across systems, reflecting interaction intensities. In complexes CNT+OH\textsubscript{20}+G5 and CNT+OH\textsubscript{25}+G5, $\rho$ values exceed 0.035~a.u., indicating strong covalent-like interactions, possibly associated with highly directional and stable hydrogen bonding. In contrast, systems like CNT+G2 or CNT+OH\textsubscript{5}+G1 exhibit $\rho$ values below 0.015~a.u., characterizing weaker interactions dominated by van der Waals forces or dispersive hydrogen bonding, consistent with criteria established by Bader~\cite{36_bader_atoms_1994}.

\begin{figure}[tbph]
\centering
\includegraphics[width=10cm]{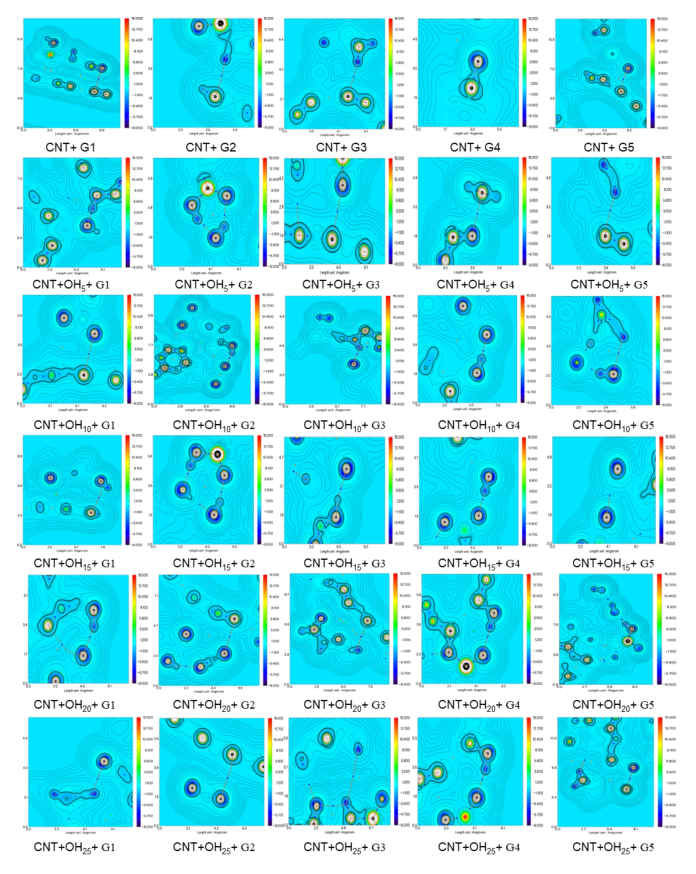}
\caption{\label{Fig:Laplacian} 2D representation of the electronic density Laplacian, $\nabla^2 \rho$.}
\end{figure}

In the analyzed systems, complexes with higher functionalization degrees like CNT\-+OH\textsubscript{20}\-+G4/G5 and CNT+OH\textsubscript{25}+G5 demonstrated Laplacian of electronic density $\nabla^2\rho$ values that remain positive but lower than in other systems. This indicates that despite significant total electronic density ($\rho > 0.10$~a.u.), local depletion exists, consistent with hydrogen bonding characteristics. In this context, Espinosa, Molins, and Lecomte~\cite{67_espinosa_hydrogen_1998} assert that systems with $\nabla^2\rho > 0$ and medium to high average $\rho$ can still exhibit hydrogen bonds with partially covalent characteristics, depending on local potential energy and kinetic energy density values. The electronic density Laplacian is depicted in Figure~\ref{Fig:Laplacian} for all systems.

Additionally, systems with less deprotonated glyphosate forms G1 and G2 display higher and more positive $\nabla^2\rho$ with lower electronic density, typical of purely electrostatic and non-directional interactions, coherent with these systems' reduced polarity. Moreover, the reduced $\nabla^2\rho$ values in highly OH-functionalized systems with more deprotonated glyphosate suggest increased local polarization and orbital reorganization, aligning with previous study associating this behavior with efficient and selective adsorption~\cite{68_bouhara_comparative_2021}.

\begin{figure}[tbph]
\centering
\includegraphics[width=10cm]{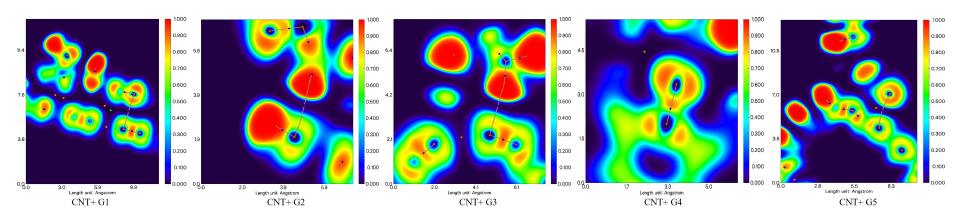}
\caption{\label{Fig:ELF} 2D representation of the electronic localization function (ELF).}
\end{figure}

The Electron Localization Function (ELF) reinforces inferences about pesticide molecule-nanotube surface bonding nature by providing an electron localization measure. ELF values exceeding 0.65, observed in CNT+OH\textsubscript{20}+G5 and CNT+OH\textsubscript{25}+G5 systems, indicate regions of high strongly localized electron pair concentrations, typical of covalent-like bonds or strongly directional donor-acceptor interactions (see Figure~\ref{Fig:ELF}). In this context higher ELF values suggest more localized electrons~\cite{69_becke_simple_1990}, indicative of stronger covalent-character bonds that promote more stable pollutant retention. Conversely, systems like CNT+OH\textsubscript{5}+G2 and CNT+G3 with ELF below 0.50 demonstrate a lower probability of localized electronic sharing, suggesting predominant electrostatic and dispersion interactions.

\begin{figure}[tbph]
\centering
\includegraphics[width=10cm]{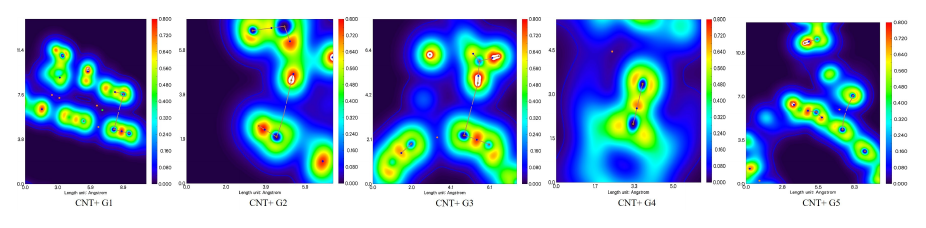}
\caption{\label{Fig:LOL} 2D representation of the electronic local orbital locator (LOL) function.}
\end{figure}

This difference is further reinforced by Local Orbital Locator (LOL) values, shown in Figure~\ref{Fig:LOL}, which help identify orbital density accumulation regions between fragments. Systems with LOL values superior to 0.60 exhibit interactions with greater electronic sharing content, contributing to enhanced complex stability and reduced glyphosate desorption propensity from the adsorbent surface~\cite{46_dong_hydrogen_2024}.

\subsection{Molecular Dynamics}
\label{MolDyn}
\begin{figure}[tbph]
\centering
\includegraphics[width=10cm]{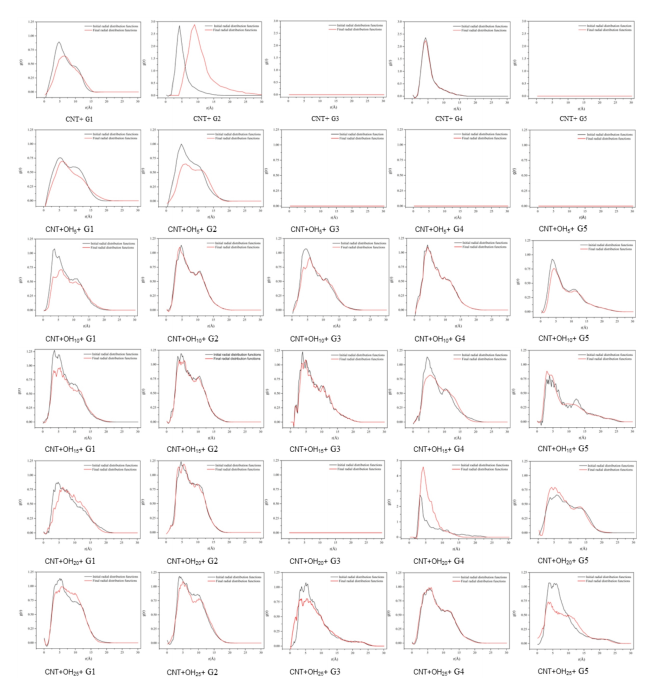}
\includegraphics[width=15cm]{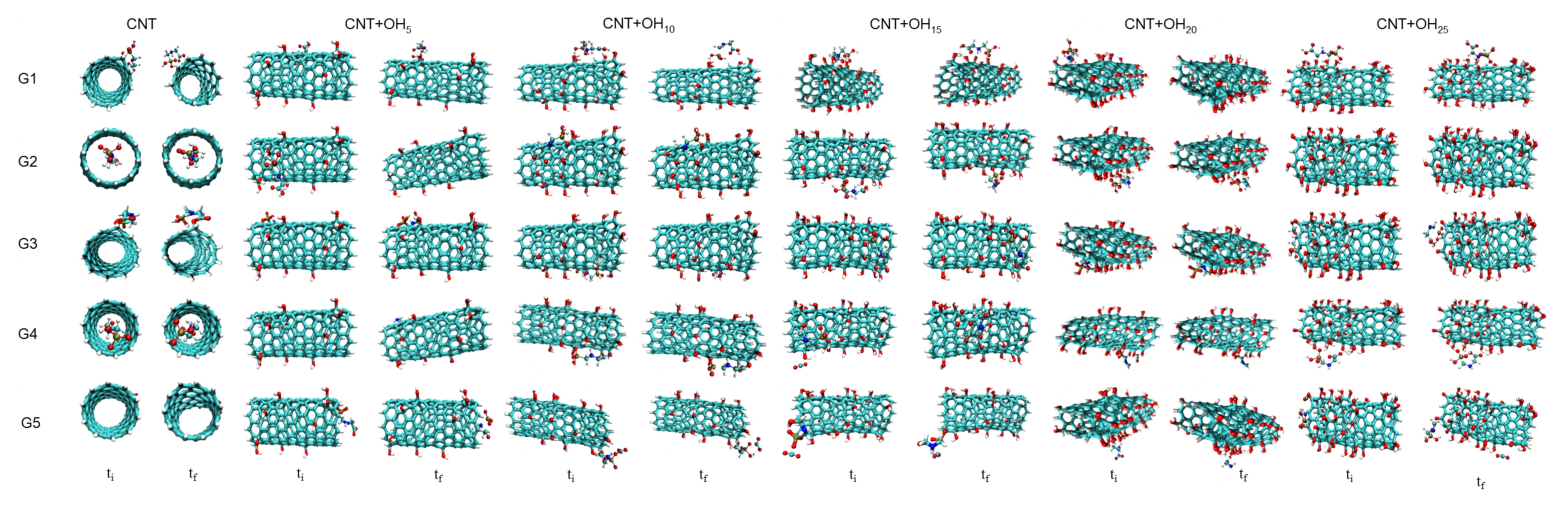}
\caption{\label{Fig:RDF} (A) Initial and final radial distribution functions calculated with VMD software~\cite{71_humphrey_vmd_1996}. (B) Initial ($t_i = 0~ps$) and final ($t_f = 100~ps$) complex configurations.}
\end{figure}

Calculating the distance between atoms or functional groups can help configure interaction types. After molecular dynamics (MD) simulations on optimized geometries, radial distribution (RDF) of glyphosate relative to nanotubes was calculated, enabling spatial organization assessment between glyphosate molecule atoms and carbon nanotube surfaces~\cite{70_Mousavi-2023}. Figure~\ref{Fig:RDF} presents RDF analyses of initial and final configurations calculated using VMD software~\cite{71_humphrey_vmd_1996}. These initial distributions exhibit Gaussian profiles, reflecting the stochastic molecular positioning algorithm employed~\cite{38_martinez_packing_2003}.

For pure CNT systems, CNT+G1 complexes demonstrated an initial peak between 4-7~{\AA}, with attenuation in the final curve, suggesting glyphosate molecule rearrangement. CNT+G3 and CNT+G4 maintained stable bond interactions with covalent or partially covalent characteristics, confirmed by electronic density, ELF, LOL, and density Laplacian data. For CNT+G2, greater glyphosate mobility around the nanotube structure was observed without establishing a definitive interaction, indicating weak interactions or localized adsorption absence, corroborating topological analysis findings. The CNT+G5 system showed no significant interaction evidence in topological and RDF analyses, suggesting adsorption absence.

With hydroxyl group introduction on nanotube surfaces, increased density and peak intensity in RDF curves was observed, attributable to greater affinity between glyphosate polar groups and nanotube functional sites, favoring hydrogen bonding and electrostatic interactions. In this context, as RDF peak intensity increases, adsorbent existence probability rises, and adsorbent-adsorbate contact becomes stronger~\cite{70_Mousavi-2023}. Additionally, authors highlight that smaller atomic distances indicate stronger molecular interactions. Peaks below 3.5~{\AA} associate with chemical or hydrogen bonding, while peaks above 3.5~{\AA} correspond to non-bonded interactions like Coulombic and van der Waals forces.

For 15, 20, and 25\% OH concentrations, increased RDF quantities were observed, indicating stable organization between glyphosate and nanotubes. In CNT+OH\textsubscript{x}+G3, G4, and G5 systems, multiple peak patterns emerged, reinforcing that functionalization promotes synergistic interactions between localized charges, resulting in enhanced herbicide immobilization and effective adsorption~\cite{73_poorsargol_role_2020,72_arabian_molecular_2021}. Moreover, with increasing hydroxyl group concentrations, contact region amplitudes between glyphosate and CNT expand, forming more distributed interactions. This indicates functionalization promotes higher selective adsorption efficiency for charged species, as OH groups elevate nanotubes' surface polarity.

\section{CONCLUSIONS}
\label{Concl}
The study results demonstrate that carbon nanotube functionalization with hydroxyl groups at 5, 10, 15, 20, and 25\% concentrations can promote electronic and structural property alterations in the systems. These modifications cause changes in system geometry and glyphosate molecule interactions with carbon nanotubes, with more deprotonated G4 and G5 forms tending to induce more pronounced rearrangements in CNT orientation and curvature.

Regarding binding energy analysis, complex stability was observed to vary based on hydroxyl concentration and glyphosate ionization degree. Higher OH concentrations and presence of more deprotonated forms (G4 and G5) exhibit greater stability with stronger interactions, potentially increasing adsorption process efficiency while simultaneously reducing adsorbent regeneration and nanotube reuse possibilities. Conversely, moderate interactions were noted in systems like CNT+OH\textsubscript{x}+G1 and CNT+OH\textsubscript{x}+G3, suggesting system equilibrium and stability that are environmentally and economically viable, enabling nanotube reutilization.

Frontier orbital and electronic coupling integral ($J_{oc}$, $J_{un}$, $J$) analyses confirmed that charge reactivity and transport were optimized in systems with 20-25\% OH in the presence of G4 and G5 glyphosate forms. Additionally, topological characterization studies verified donor-acceptor interactions with strong covalent contributions, as indicated by electronic density values and their Laplacians. Radial distribution function (RDF) profiles demonstrated that functionalization and glyphosate molecular state favor spatial organization on nanotube surfaces, increasing contact regions and reducing molecular mobility.

Therefore, OH-functionalized carbon nanotubes show promise in developing materials for glyphosate molecule detection and capture, regardless of their ionized state in the environment, with potential applications in environmental monitoring and remediation. Future studies are recommended to analyze economic feasibility and practical large-scale nanomaterial production.

\section*{CRediT authorship contribution statement}
\textbf{H.~T.~Silva}: Formal analysis, Investigation, Writing-original draft, Writing-review \& editing.
\\
\textbf{L.~C.~S.~Faria}: Formal analysis, Investigation, Writing-original draft, Writing-review \& editing.
\\
\textbf{T.~A.~Aversi-Ferreira}: Formal analysis, Investigation, Writing-original draft, Writing-review \& editing.
\\
\textbf{I.~Camps}: Conceptualization, Formal analysis, Methodology, Project administration, Resources, Software, Supervision, Writing-review \& editing.

\section*{Declaration of Competing Interest}

The authors declare that they have no known competing financial interests or personal relationships that could have appeared to influence the work reported in this paper.

\section*{Data Availability}
\label{data_avail}
The raw data required to reproduce these findings are available to download from Zenodo repository.

\section*{Acknowledgements}
I.C. is grateful to the Brazilian funding agency CNPq for the research scholarship (304937\allowbreak/2023-1). This study was financed in part by the Coordena\c{c}\~ao de Aperfei\c{c}oamento de Pessoal de N\'{\i}vel Superior-Brasil (CAPES)-Finance code 001. Part of the results presented here were developed with the help of CENAPAD-SP (Centro Nacional de Processamento de Alto Desempenho em S\~ao Paulo) grant UNICAMP/FINEP-MCT, and the National Laboratory for Scientific Computing (LNCC/MCTI, Brazil) for providing HPC resources of the Santos Dumont supercomputer and CENAPAD-UFC (Centro Nacional de Processamento de Alto Desempenho, at Universidade Federal do Cear\'a).

\newpage

\end{document}